\documentclass[reprint,showpacs,preprintnumbers,amsmath,amssymb,superscriptaddress, aps, pra, nofootinbib]{revtex4-1}
\eprint
\usepackage{lipsum}
\usepackage[symbol]{footmisc}
\usepackage{graphicx}
\usepackage{tabularx}
\usepackage{bm}
\usepackage{color}
\usepackage{esvect}
 
\usepackage{ulem}

\usepackage{comment}

\usepackage{ulem}

\begin{document}
	
	\title{Probing Electron-Hole Coherence in Strongly-Driven Solids}
	
	\author{Christian Heide} 
	\thanks{These authors contributed equally.}
	\affiliation{Stanford PULSE Institute, SLAC National Accelerator Laboratory, Menlo Park, CA 94025, USA}
	\affiliation{Department of Applied Physics, Stanford University, Stanford, CA 94305, USA}
	\author{Yuki Kobayashi}
	\thanks{These authors contributed equally.}
	\affiliation{Stanford PULSE Institute, SLAC National Accelerator Laboratory, Menlo Park, CA 94025, USA}
	\affiliation{Department of Applied Physics, Stanford University, Stanford, CA 94305, USA}
	\author{Amalya Johnson}
	\affiliation{Department of Materials Science and Engineering, Stanford University, Stanford, CA 94305, USA} 
	\author{Fang Liu}
	\affiliation{Stanford PULSE Institute, SLAC National Accelerator Laboratory, Menlo Park, CA 94025, USA}
	\affiliation{Department of Chemistry, Stanford University, Stanford, CA 94305, USA}
	\author{Tony F. Heinz}
	\affiliation{Stanford PULSE Institute, SLAC National Accelerator Laboratory, Menlo Park, CA 94025, USA}
	\affiliation{Department of Applied Physics, Stanford University, Stanford, CA 94305, USA}
	\author{David A. Reis}
	\affiliation{Stanford PULSE Institute, SLAC National Accelerator Laboratory, Menlo Park, CA 94025, USA}
	\affiliation{Department of Applied Physics, Stanford University, Stanford, CA 94305, USA}
	\author{Shambhu Ghimire}
	\affiliation{Stanford PULSE Institute, SLAC National Accelerator Laboratory, Menlo Park, CA 94025, USA}
	
	\date{\today}
	
	\maketitle 
	
	\textbf{High-harmonic generation (HHG) is a coherent optical process \cite{ferray_multiple-harmonic_1988,Ghimire2011} in which the incident photon energy is up-converted to the multiples of its initial energy. In solids, under the influence of a strong laser field, electron-hole (e-h) pairs are generated and subsequently driven to high energy and momentum within a fraction of the optical cycle. These dynamics encode the band structure \cite{Ghimire2011, Luu2015, Vampa2015a, Ghimire2019}, including non-trivial topological properties \cite{Liu2017, Silva2019, Chacon2020, Mrudul2020, Bai2020, Baykusheva2021, Schmid2021} of the source material, through both intraband current and interband polarization, into the high harmonic spectrum. In the course of this process, dephasing between the driven electron and the hole can significantly reduce the HHG efficiency \cite{Floss2019, Vampa2015a}. Here, we exploit this feature and turn it into a measurement of e-h coherence in strongly driven solids. Utilizing a pre-pump pulse, we first photodope monolayer molybdenum disulfide and then examine the HHG induced by an intense infrared pulse. We observe clear suppression of the HH intensity, which becomes more pronounced with increasing order. Based on quantum simulations, we attribute this monotonic order dependence as a signature of ultrafast electron-hole dephasing, which leads to an exponential decay of the inter-band polarization, proportional to the sub-cycle excursion time of the e-h pair. Our results demonstrate the importance of many-body effects, such as density-dependent decoherence in HHG and provide a novel platform to probe electron-hole coherence in strongly driven systems.}\\
	\begin{figure*}[t!]
		\begin{center}
			\includegraphics[width=14cm]{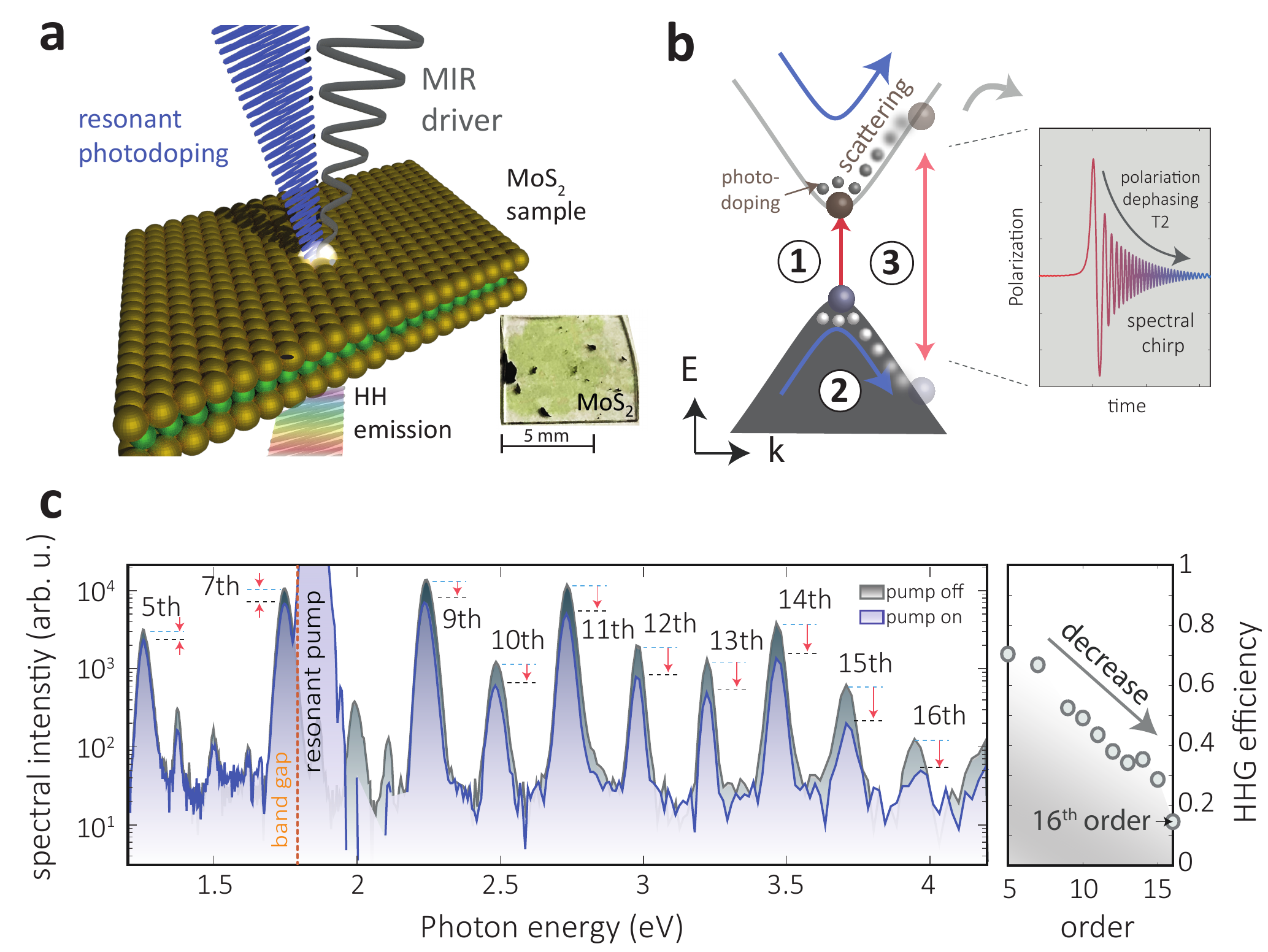}
			\caption{\textbf{Electron-hole dephasing during high harmonic generation.}
				\textbf{a}, Experimental configuration. An intense 5\,$\mu$m-MIR pulse is focused to a single-crystal MoS$_2$ sample, generating high-harmonic (HH) radiation. The HHs are collected and measured in a transmission geometry. A 660\,nm laser pulse is applied to resonantly inject photocarriers in the sample. The bottom right shows an optical microscope image of the mm-sized sample used in our experiments. . 
				\textbf{b}, Schematic illustration of the e-h dynamics near the band gap. The MIR-pulse generates e-h pairs via inter-band transitions (step 1). Within a quarter of cycle of the MIR drive field, these e-h pairs are accelerated in the band structure (step 2). The anharmonicity of the bands and the time-dependent energy spacing between e-h (polarization) give rise to HHG. In the case of photodoping, the e-h scattering rate may become enhanced, strongly suppressing higher-order harmonics. 
				In the right panel, the inter-band current (polarization response) is sketched for a quarter period of the MIR-pulse. As the e-h pair is accelerated, the frequency of the polarization response increases, giving rise to a blue chirp of the HH emission. Due to e-h decoherence, these higher-order harmonics are reduced, as parameterized by dephasing time $T_2$.
				\textbf{c}, Measured HH spectrum, ranging from 5$^\text{th}$ to 16$^\text{th}$ HH order for two cases: gray without photodoping and blue with photodoping. The red arrows compare the peak spectral intensity of each harmonic order with and without photodoping. The integrated and normalized efficiency (pump-on/pump-off) for each harmonic order is shown on the right panel.}	
			\label{fig: 1}
			\vspace{-10pt}
		\end{center}
	\end{figure*}
	Transparent solids subjected to a sufficiently strong laser field produce high harmonics, with photon energies that can greatly exceed their band-gap energy \cite{Ghimire2011}. The underlying microscopic dynamics has been studied intensively during the past decade \cite{Ghimire2019}. Two main channels have been identified and their respective roles were examined \cite{Vampa2015,Luu2015, Liu2017, Yoshikawa2019, Ghimire2019}. First, high harmonics (HHs) can be generated from the driven nonlinear intra-band current, associated with non-parabolic bands \cite{Luu2015, Ghimire2011a}. Second, HHs can arise from the dipole-response of the e-h pairs \cite{Vampa2015, Wu2017}, which produce an interband polarization. The two sources are shown in Fig.~\ref{fig: 1}\textbf{b} with the blue and red arrows, respectively. 
	The unique feature of the interband channel is the particular sensitivity to coherence between the excited electrons and their associated hole, which is different from common emission radiative processes, such as fluorescence emission \cite{Mak2010}. 
	A particular feature of solid-state systems that contrasts with the well-studied response in gases is that the above-mentioned coherence between the driven e-h pairs may be readily disrupted, either through intrinsic mechanisms, such as electron-phonon scattering, or though extrinsic mechanisms, such as electron scattering with crystal defects, impurities, or photogenerated charge carriers. These processes induce e-h decoherence and thus impede inter-band HHG \cite{Bigot1991, Vampa2014, Floss2019}.\\
	Measurements such as coherent spectroscopy, photocurrent and streaking indicate that the e-h dephasing time in solids can range from nanoseconds down to attoseconds, depending on the carrier concentration and their driving parameters \cite{Bigot1991, Petek1997, Cundiff2016, Seiffert2017, Higuchi2017, Heide2021a}. To date, however, sensitive experimental probes of e-h coherence under conditions of strong driving electric fields, as are crucial for HHG, have yet to be developed.
	
	Here we intentionally disrupt the electron-hole coherence through photodoping and exploit the underlying dynamics of the HHG process to probe the coherence lifetime in a controlled manner, with sub-cycle temporal resolution. As shown schematically in Fig.~\ref{fig: 1}\textbf{a}, we use two laser pulses: a strong mid-infrared (MIR) that drives the HHG process and a relatively weak visible pump pulse that perturbs the HHG process by resonantly injecting charge carriers from the valence to the conduction band.\\
	
	In our experiment we use monolayer molybdenum disulfide (MoS$_2$, direct bandgap: 1.8\,eV) as a model system to study the role of electron-hole dephasing in the HHG process. MoS$_2$ serves as an ideal material system as it has a reduced dielectric screening and strong many-body Coulomb interactions \cite{Mak2010}. Furthermore, it has strong dipole coupling and a band gap in the visible range, which results in high harmonics ranging from 1\,eV to 4\,eV when pumped with mid-infrared pulses having field strengths on the order of 1\,V/nm \cite{Liu2017, Yoshikawa2019}.\\
	A representative HH spectrum for the 5\,$\mu$m drive field (cycle period of $T = 16.7\,$fs) with and without the 660\,nm pump pulse is shown in Fig.~\ref{fig: 1}\textbf{c}. For a peak electric field strength of 0.6\,V/nm (intensity 7$\times10^{10}$\,W/cm$^2$), we measure even- and odd-order harmonics, ranging from 5$^\text{th}$ to 16$^\text{th}$ harmonic order. The fluence $F$ of the resonant pump at 660\,nm with a pulse duration of about 100\,fs is chosen below the damage threshold of MoS$_2$. The delay the MIR pulses with respect to the visible pulses is set to $\Delta t$ = 1\,ps, to avoid their temporal overlap (see Methods). The observation of ten distinct harmonic orders enables us to obtain quantitative information about the response of different harmonics to various photodoping conditions. \\ 
	When the resonant pump pulse is applied, we observe that all measured harmonics are significantly suppressed, with stronger suppression of the higher harmonic orders. The right panel shows the integrated spectral intensity as a function of the HH order, normalized to the condition without photodoping. Whereas the 5$^\text{th}$ order harmonic is reduced by 30\%, higher orders are suppressed up to 85\% for the 16$^\text{th}$ harmonic order. \\
	
\begin{figure}[t!]
		\begin{center}
			\includegraphics[width=9cm]{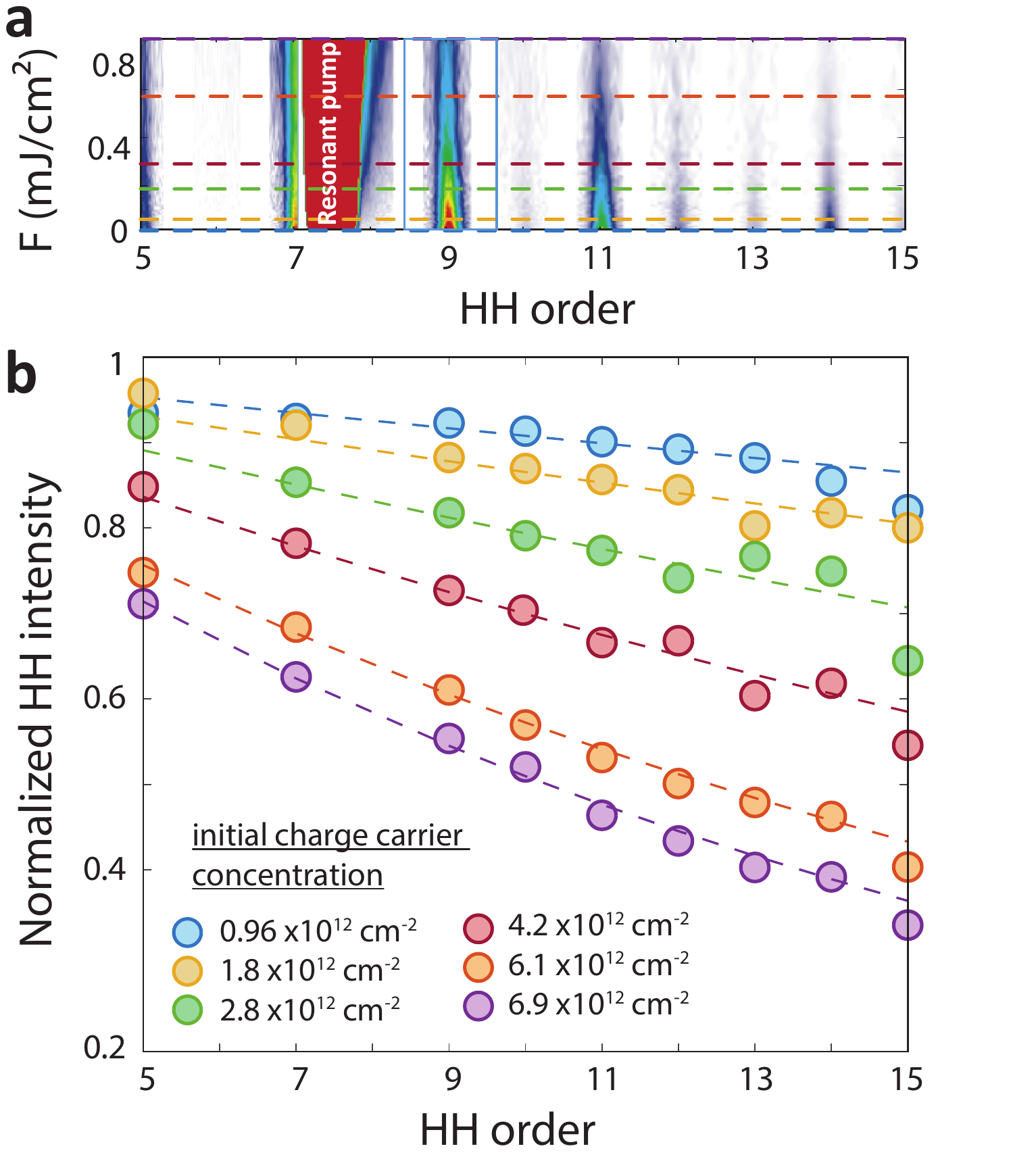}
			\caption{\textbf{Systematic dependence of the high harmonics on photodoping.} \textbf{a}, Measured HH spectrum as a function of the pump fluence $F$. Line-outs (dashed lines) for six different pump fluences are plotted in (b). \textbf{b}, The normalized HH intensity as a function of HH order for different initial carrier densities at $\Delta t = 1$\,ps. Higher-order harmonics are more strongly affected by an increase in $n$. The solid lines are fit functions, taking dephasing time $T_2$ of the electron-hole pair into account (Eq.\,\eqref{eq. 1}, outlined in Methods).}	
			\label{fig: 2}
			\vspace{-10pt}
		\end{center}
	\end{figure}

	To study systematically the dependence of the high harmonics, we focus on the variation of the initial photocarrier concentration $n$ and measure the corresponding HH spectrum from the delayed non-resonant strong-field drive. To obtain $n$, we first determine the absorbance as a function of pump fluence and assume that each absorbed photon generates an electron-hole pair. We then calculate $n$ at $\Delta t$ = 1\,ps, by taking rapid exciton-exciton annihilation into account (see \cite{Sun2014} and Methods). Figure~\ref{fig: 2}\textbf{a} shows the obtained HH spectrum as a function of pump fluence. We observe that all measured HH orders decrease with increasing $n$. To compare different orders to one another, we plot the integrated and normalized spectral intensity in Fig.~\ref{fig: 2}\textbf{b}. For $n$ = $0.96\times 10^{12}$ cm$^{-2}$ (blue circles) the HH intensity is reduced by 8\% for the 5$^\text{th}$ harmonic and decreases almost monotonically to 15\% for the highest observed harmonic. Increasing the photodoping to $n$ = $6.9\times10^{12}$ cm$^{-2}$ (purple circles) results in a decrease of 30\% for the 5$^\text{th}$ harmonic up to 65\% for the 15$^\text{th}$ harmonic.
	
	\begin{figure*}[t!]
		\begin{center}
			\includegraphics[width=17cm]{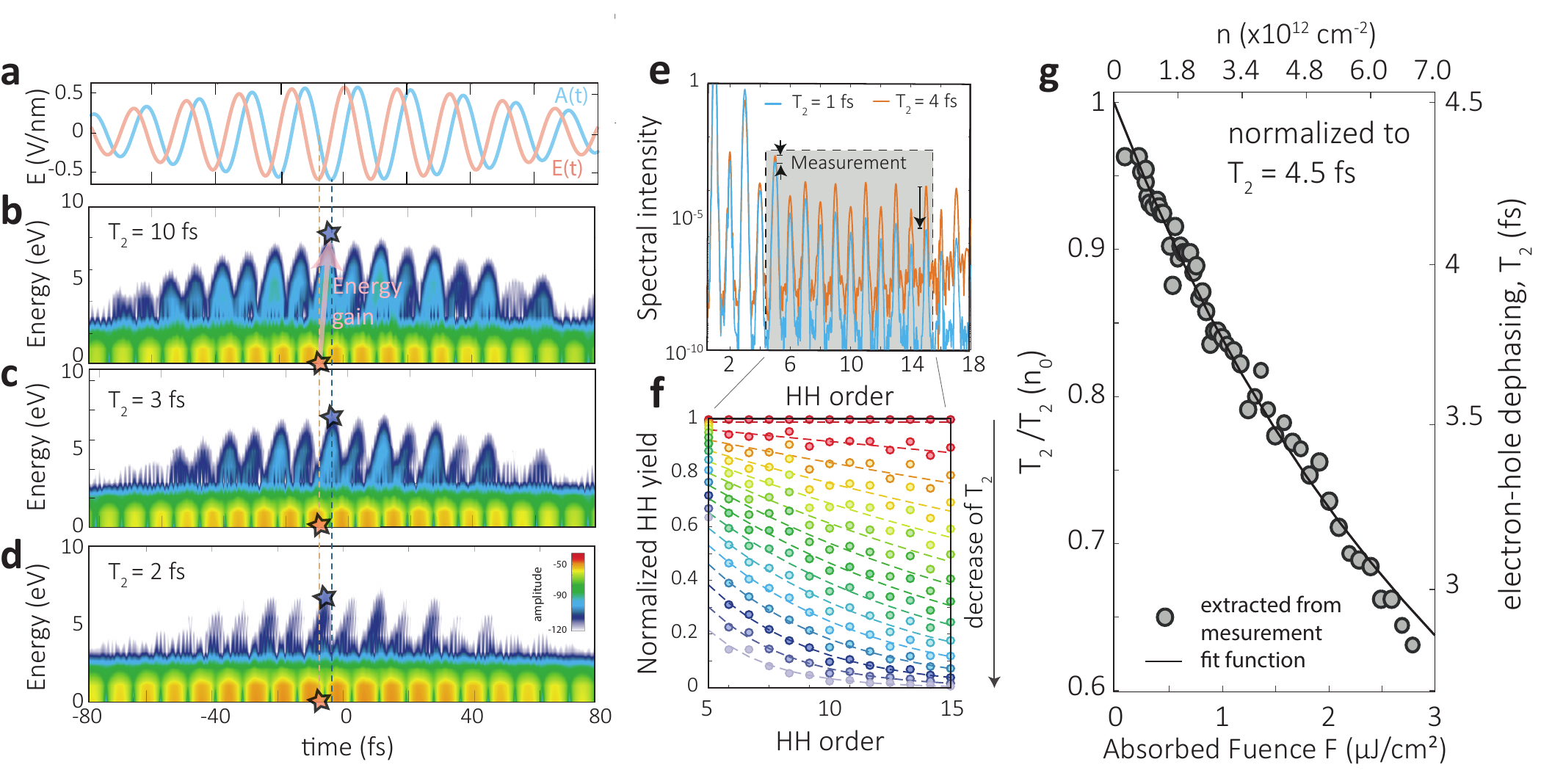}
			\caption{\textbf{Numerical simulations results and comparison to experimental data.}
				\textbf{a--d}, Time-frequency analysis of the HHG process for $T_2 = $10\,fs in (b), 3\,fs in (c) and 2\,fs in (d). When the electron is driven by the vector potential $A(t)$, the electron gains energy up to the point, where $A(t)$ reaches its greatest amplitude. When the dephasing time is longer than a quarter of the optical cycle, the e-h pair maintains its coherence during this process, resulting in the emission of high-energy photons, as shown in (b). The brown star indicates the time of the e-h generation, which yields the highest photon energy, indicated with the blue star. For $T_2$ = 1\,fs and 4\,fs the higher-order harmonics arising from long trajectories are suppressed. 
				\textbf{e}, HH spectrum for different dephasing times. Reducing the dephasing time results in a suppression of the spectral intensity for high harmonic orders. The spectral range measured in Fig.~\ref{fig: 2} is boxed. 
				\textbf{f}, Integrated and normalized HHG yield as function of HH order. The circles are obtained from the SBE simulations and the dashed lines are fit functions using Eq.\,\eqref{eq. 1}, outlined in the Method section. Each color is represented by one characteristic $T_2$ value ranging from 4.25\,fs (red) to 0.75\,fs (gray) with a step size of 0.25\,fs. 
				\textbf{g}, Fitting the numerical results, shown in (f) to the experimental HH spectrum (Fig.~\ref{fig: 2}) allows us to determine the relative change of $T_2$ as a function of absorbed pump fluence or photocarrier density $n$. The solid line is based on a model function assuming a free electron gas, i.e., $T_2^{-1} \propto n$ (see Methods).}
			\label{fig: 3}
			\vspace{-10pt}
		\end{center}
	\end{figure*}
	
	To investigate the role of photodoping in the HHG process, we begin by reexamining the details of the microscopic mechanism for solid-state HHG, illustrated in Fig.~\ref{fig: 1}\textbf{b}. Under the presence of the laser field, electron-hole pairs are mostly generated near the band gap at the $K$ and $K'$ points (step 1) where the dipole coupling between valence and conduction bands is maximized. Subsequently, the field transiently drives them to high energy and momentum states (step 2), giving rise to coupled intra-band motion (current, blue arrow), inter-band transitions (polarization, red arrows), and scattering events. The right panel in Fig.~\ref{fig: 1}\textbf{b} shows schematically the inter-band polarization response during a quarter of an optical cycle. First, a polarization oscillating at low frequencies is generated, given by the band-gap energy, followed by high frequencies as the electron-hole pair is subsequently driven to higher energies. This polarization response is the dominant source for harmonics around and above the band gap (see Methods), with an emission time defined by the recombination step in the semi-classical picture (step 3) \cite{Vampa2015a}. In this scheme, we note that harmonics originating from longer e-h trajectories are more sensitive to decoherence as the corresponding excursion time allows for more scattering events. By measuring the order-dependent variation in the harmonic intensity, we can infer the lifetime of the electron-hole coherence.\\
	
	In order to extract a dephasing time from the order dependent reduction in the HHG intensity, we simulate HHG using a tight-binding model Hamiltonian with a hexagonal graphene-like band structure with two different sub-lattices. Such a Hamiltonian exhibits broken inversion symmetry, leading to the production of both odd- and even-order harmonics \cite{Motlagh2019}. The electron dynamics is determined by solving numerically the semiconductor Bloch equations (SBE) in the presence of the laser field, using the Houston representation \cite{Li2019a}. Importantly, this formalism allows us to introduce phenomenologically an e-h dephasing time constant $T_2$ (see Methods and \cite{Floss2019}). We note that the parameter $T_2$ accounts for both electron-electron and electron-phonon scattering as decoherence processes leading to the decay of the e-h polarization.
	
	The results of the simulations are first analyzed by plotting time-frequency spectrograms, which provide an intuitive visualization of the dynamics of the electron-hole trajectories. Figures~\ref{fig: 3}\textbf{a}--\textbf{d} show the results for the dephasing times of $T_2 = 2, 3, 10$\,fs. For our experimental parameters and particularly for harmonics above the band gap, the inter-band polarization dominates over the intra-band current, we thus limit our discussion in Fig.~\ref{fig: 3}\textbf{b}--\textbf{d} to the inter-band polarization. For a dephasing time longer than a quarter of the optical cycle, i.e., $T_2 = 10$\,fs (Fig.~\ref{fig: 3}\textbf{b}), the e-h pair maintains its coherence up to the point of maximal e-h energy separation, i.e., 8\,eV (blue star). Here, the e-h pair is generated when the electric field amplitude $|E(t)|$ has a peak (red brown) and the vector potential $|A(t)|$ has a minimum, and reaches its maximal energy-spacing after a quarter of the optical cycle, i.e., when $|A(t)|$ has its peak (blue star). Reducing the dephasing time to 3\,fs in Fig.~\ref{fig: 3}\textbf{c} and 2\,fs in Fig.~\ref{fig: 3}\textbf{c} suppress the coherence for the longest short trajectories and thus, the maximal energy achieved is exponentially reduced. We observe for $T_2 = 3$\,fs a maximal photon energy of about 7\,eV, which drops down to 6\,eV for $T_2 = 2$\,fs.\\
	Figure~\ref{fig: 3}\textbf{e} shows the total calculated HH spectrum for different $T_2$ values. Within our model a shorter dephasing time clearly suppresses higher orders more strongly.\\
	
	To evaluate the trends for suppression across the different harmonic orders, we again integrate the spectral intensity and normalize it to $T_2(n_0) = T/4$, as indicted in Fig.~\ref{fig: 3}\textbf{f}. Different colors show various dephasing times from $T_2 = 4.25$\,fs (red) to $T_2 = 0.75$\,fs (gray), with respect to $T_2(n_0)$ = 4.5\,fs. We note that the shape of the drop is also captured by a simple analytic description presented in the Method section.\\
	
	We can now compare the numerical simulation with the experimental results, shown in Fig.~\ref{fig: 2}, and relate $n$ and $T_2$. As each curve in Fig.~\ref{fig: 3}\,\textbf{f} is described by the single $T_2$ parameter, we fit the numerical results to the experimental data. Importantly, the numerical model captures the experimental data quite well with just one free parameter. This allows us to determine the relationship between $n$ and $T_2$ in Fig.~\ref{fig: 3}\,\textbf{g}. By increasing $n$, we observe that $T_2$ decreases by 40\% for the maximal photocarrier doping of 6.9$\times$10$^{12}$\,cm$^{-2}$. 
	
	Based on the simulations, we estimate an upper boundary of e-h coherence as a quarter of the optical cycle (i.e., 4.5 fs) for no or weak photodoping, which decreases to 2.5\,fs for $n = 6.9\times 10^{12}$\,cm$^{-2}$. For $T_2$ longer than a quarter of an optical cycle, the one-to-one relation between excursion times of the e-h pairs and the emitted photon energies will be lost, resulting in relatively flat or non-monotonic reduction of the high harmonic yield with increasing order.
	
	In the presented work, the effect of photodoping on the HHG process is explained as a consequence of an enhancement in the rate of e-h dephasing. Dephasing mainly affects interband-dominated harmonics around and above the band gap, as supported by our numerical simulations. In contrast, intra-band current may dominate HHG for harmonics at energies well below the band gap \cite{Vampa2014}. Assuming that scattering among the electrons does not decrease the intraband current significantly these harmonics should actually be enhanced under photodoping because of the higher carrier densities, as also suggested in \cite{Wang2017, Nagai2020}. For still higher applied pump intensities, i.e., when $n \gtrsim 10^{13}$\,cm$^{-2}$, band-gap renormalization in monolayers \cite{Steinhoff2014}, as well as state filling for peak pump fluence $F > F_\text{sat} = $0.13\,mJ/cm$^2$ might become important for the overall HH intensity. The monotonic depletion of the HH intensity as a function of harmonic order has also been observed in our auxiliary measurements of bulk MoS$_2$, indicating that strong excitonic effects and band-gap renormalization are not primarily responsible for the observed monotonic reduction of the HH intensity. 
	
	In summary, we employed an all-optical approach based on high-harmonic generation to measure the coherence of strongly driven electron-hole pairs in solid materials on sub-cycle timescales. In the experiment, we control the decay rate of the e-h coherence by photodoping monolayer MoS$_2$ and measure the corresponding HHG response. We find that an increased charge-carrier concentration reduces the overall efficiency of HHG, with more prominent effects observed as the high-harmonic order increases. Within the framework of the semiconductor Bloch equations, we attribute this observation to enhanced dephasing, which produces an exponential decay of the inter-band polarization that scales with the excursion time of the e-h pair. Our results highlight the importance of many-body effects, such as density-dependent coherence in HHG, and advance understanding of coherence in solid-state HHG, with ramifications for compact all-optical solid-state spectroscopy, as well as application such as short-wavelength light sources and attosecond pulse generation.

	\section{Acknowledgments}
	We thank Giulio Vampa, Ignacio Franco, Azar Oliaei Motlagh and Hamed Koochaki Kelardeh for fruitful discussions. This work was supported by	the US Department of Energy, Office of Science, Basic Energy Sciences, Chemical Sciences, Geosciences, and Biosciences Division through the AMOS program. F.L. acknowledges support from a Terman Fellowship and startup funds from the Department of Chemistry at Stanford University. Y.K. acknowledges support from the Urbanek-Chorodow Fellowship from Stanford University and C.H. from the Humboldt Fellowship and the W. M. Keck Foundation.

	\section{Author contributions}
	C.H., Y.K. and S.G. conceived the study. A.J. and F.L. provided the high-quality MoS$_2$ samples. C.H. and Y.K. performed the measurements, evaluated the data and performed the numerical simulations. T.F.H., D.A.R. and S.G. supervised the work. All authors discussed the results.

	%

	\setcounter{equation}{0}
	\setcounter{figure}{0}
	\setcounter{table}{0}
	\makeatletter
	\renewcommand{\theequation}{M\arabic{equation}}
	\renewcommand{\thefigure}{M\arabic{figure}}
	
	\section{Methods}
	\subsection{Experimental setup}
	
\begin{figure}[h!]
		\begin{center}
			\includegraphics[width=8cm]{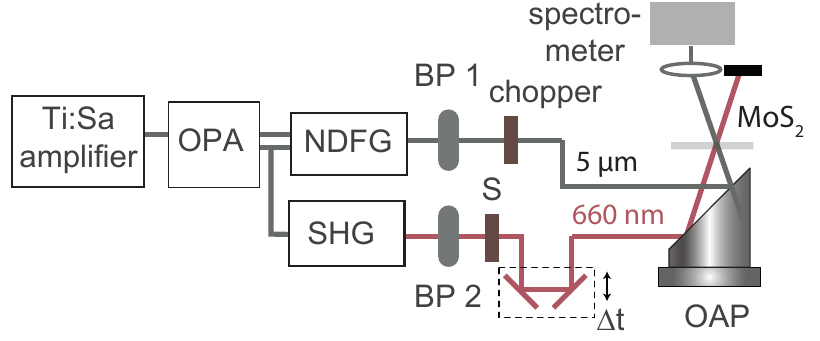}
			\caption{Schematic illustration of the experimental setup for the HH measurement.}	
			\label{fig: S4}
			\vspace{-10pt}
		\end{center}
	\end{figure}

	The experimental setup is sketched in Fig.~\ref{fig: S4}. An amplified Titanium:sapphire laser system (Evolution, Coherent Inc., 6\,mJ, 45\,fs, 790\,nm, 1kHz) is used to pump an optical parametric amplifier (OPA, TOPAS-HE, Light Conversion Inc.). The signal ($\sim1300$\,nm) and idler ($\sim1900$\,nm) from the OPA are mixed in a GaSe crystal (Eksma Optics Inc., z-cut, 0.5\,mm thick) for difference-frequency generation. The resulting mid-infrared radiation is cleaned by a bandpass filter (BP 1), centered at 5.0\,$\mu$m (Thorlabs Inc., FB5000-500). The polarization of the MIR beam is controlled by a zero-order MgF$_2$ half-wave plate. Part of the signal is frequency doubled in a BBO crystal to obtain the resonant pre-pump pulse. A bandpass filter (BP 2) centered at 660\,nm (Thorlabs, FB660-10) is used to clean up the spectrum. Both pulses are focused by a silver $90^\circ$ silver off-axis parabolic mirror (OAP) with a focal length of 100\,mm with a small angle to each other to the sample. The spot sizes are 100\,$\mu$m ($1/e^2$ intensity radius) for the MIR beam and 120\,$\mu$m for the resonant pump beam. A 20$\times$ Mitutoyo microscope objective attached to a CCD camera can be placed directly behind the sample to characterize the spot size at the focal plane of the 660\,nm beam and align the spot on the monolayer sample. The spot size of the MIR-beam is characterized with a beam profiler (Dataray, WinCamD-IR-BB). The generated high harmonics are collected and focused by CaF$_2$ lenses and directed into a spectrometer equipped with a thermoelectronically cooled CCD camera (Princeton Instruments Inc., Pixis 400B). All HH spectra were measured in a transmission geometry and under ambient conditions. The sample orientation for the linearly polarized MIR laser field was aligned to maximize even-order harmonics, i.e., along the $\Gamma - M$ direction of the crystal \cite{Liu2017}.\\
	
	The monolayer MoS$_2$ was prepared from a single crystal (SPI supplies) using gold-tape exfoliation, which yields high-quality mm-sized single-crystalline flakes \cite{Liu2020}, and was supported on a transparent fused silica substrate. 
	
	For the experimental data presented in the main part of this paper, we fixed the delay time at $\Delta t =$ 1\,ps, so that the two 100\,fs pulses do not overlap temporally. There are two approaches to control the density of photoinjected carriers: 1) by fixing the time delay and controlling the pump fluence of resonant pulse (as discussed in the main text) or 2) by changing the time delay between the resonant pump-pulse and the MIR pulse using a mechanical translation stage. Figure~\ref{fig: S4}\textbf{a} and \textbf{b} show the HHG-yield as a function of $\Delta t$. The pump fluence was fixed at 0.6\,mJ/cm$^2$. We observe that for $\Delta t < 0$, the HHG yield does not depend on the pump pulse, whereas for $\Delta t = 0$, maximal depletion is obtained. For $\Delta t > 0$, the signal recovers on a picosecond time scale as the carriers density decreases. For the maximal delay of $100$\,ps, the HH signal has almost entirely recovered.\\ 
	Figure~\ref{fig: S4}\textbf{c} shows the integrated spectral intensity for the 9$^\text{th}$ harmonic. For each time step, a mechanical chopper is used to obtain the ratio of the HH spectrum for pump-on and pump-off, which allows us to normalize the reduction of the HH intensity at each time step. We can fit the data to a bi-exponential decay for $\Delta t > 0$. For the 9th harmonic, we obtain $T_\text{short} = 2\pm 0.3$\,ps and $T_\text{long} = 140\pm 10$\,ps. By comparing these values to the literature, we associate the first time constants to relaxation of excited carriers and exciton-exciton annihilation and the second to radiative recombination \cite{Nie2014, Aleithan2016}. 
	
	\begin{figure}[t!]
		\begin{center}
			\includegraphics[width=8cm]{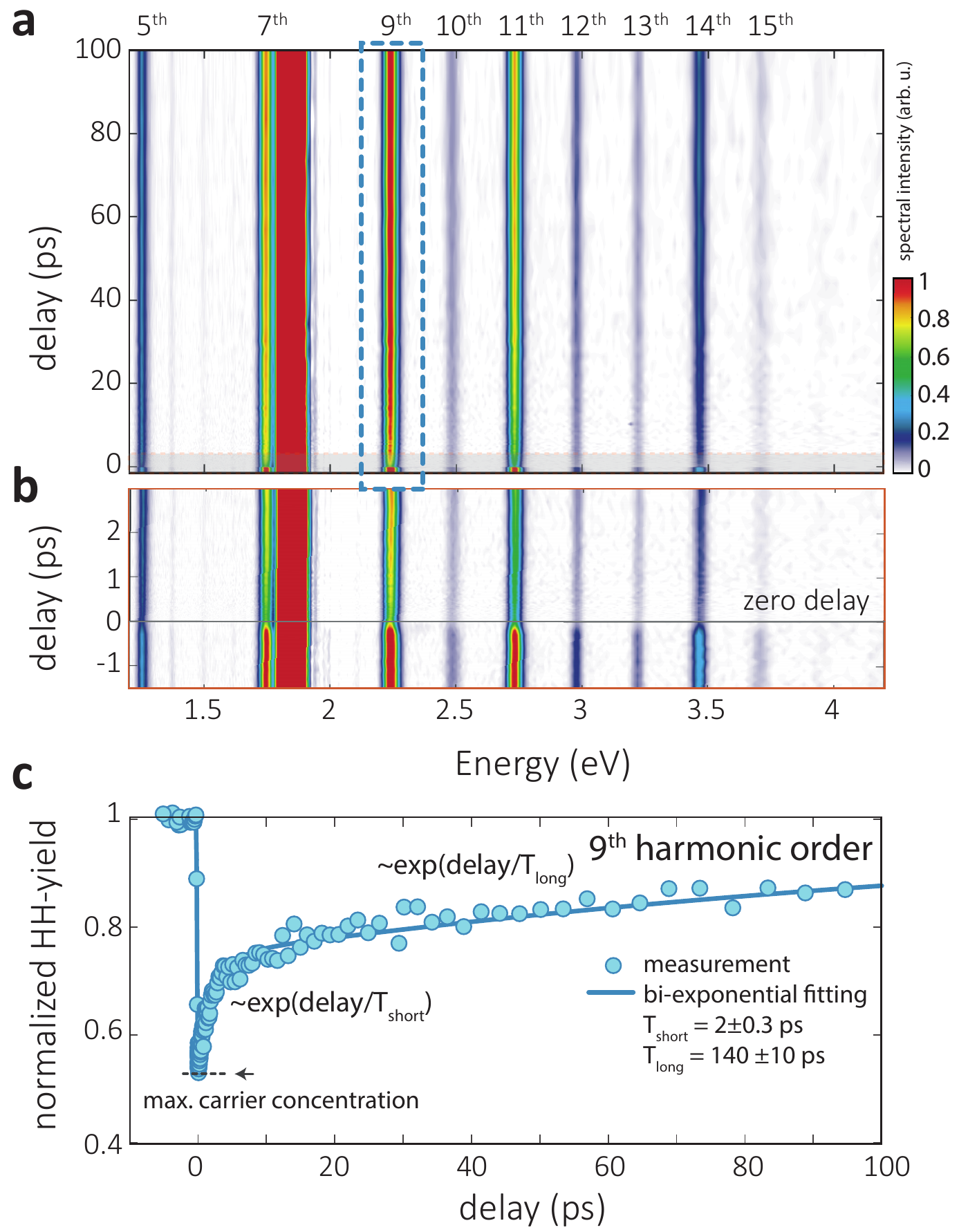}
			\caption{\textbf{Delay dependence.} 
				\textbf{a}, HH spectrum of monolayer MoS$_2$ as a function of the time delay for $\Delta t$ = -1\,ps to $\Delta t$ = 100\,ps. For $\Delta t < 0$, the MIR-pulses arrive first, and all measured harmonics are maximized. For $\Delta t = 0$, the concentration of photocarriers is maximized, and all harmonics are suppressed. For $\Delta t > 0$, the signals recover on a picosecond timescale as the carrier concentration decreases. 
				\textbf{b}, Magnified response around $\Delta t = 0$.
				\textbf{c}, Integrated spectral intensity of the 9$^\text{th}$ harmonic. The HHG recovery exhibits two time scales: $T_\text{short} 
				\approx 2$\,ps, indicating relaxation of excited carriers as well as exciton-exciton interaction and $T_\text{long} \approx 
				140$\,ps (radiative recombination).}	
			\label{fig: S2}
			\vspace{-10pt}
		\end{center}
	\end{figure}
	
	\subsection{Absorbance and charge carrier concentration}
\begin{figure}[h!]
		\begin{center}
			\includegraphics[width=7cm]{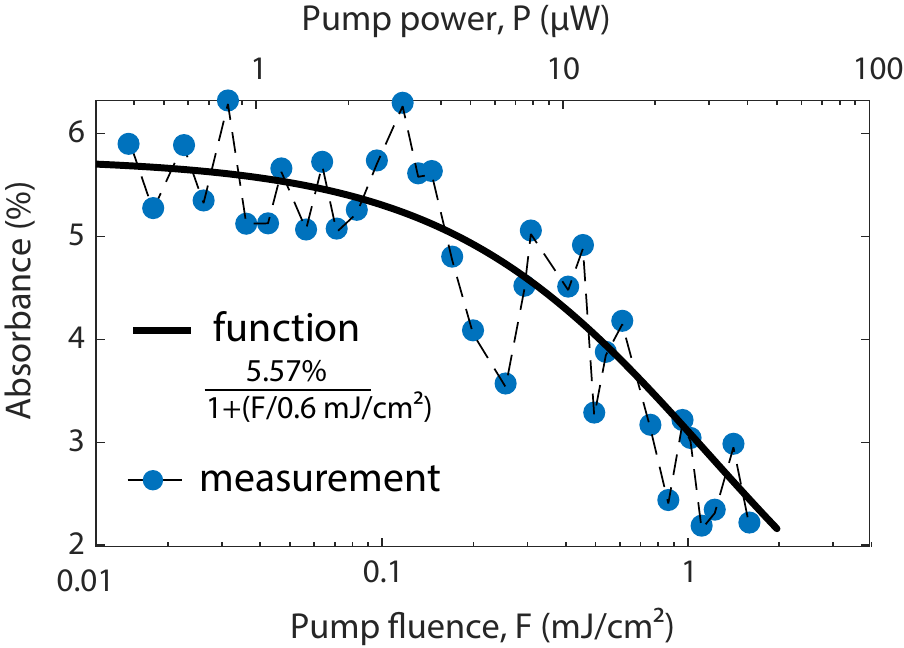}
			\caption{\textbf{Absorbance measurement in monolayer MoS$_2$.} For a low pump fluence, the absorbance is determined to 5.57\%. At a pump fluence of $F_\text{sat}$ = 0.6\,mJ/cm$^2$, the absorption becomes saturated.}	
			\label{fig: S5}
			\vspace{-10pt}
		\end{center}
	\end{figure}

	To estimate the density of photocarriers $n$ in the Mo$S_2$ monolayer, we directly measure the absorbance of the sample as a function of the incident laser fluence, as shown in Fig.\,\ref{fig: S5}. We deduce an absorbance of 5.57\% for low laser fluence, which falls to about 2\% at high pump fluences. The saturation of absorption is found at 0.6\,mJ/cm$^2$ and is larger than the values used for the HHG measurements. We assume that the initial photocarrier density $n_0$ (around $\Delta t = 0$) corresponds to each absorbed photon generating one electron-hole pair. 
	
	The charge carrier density at 1\,ps is estimated by taking rapid exciton-excition annihilation in monolayer MoS$_2$ into account. In \cite{Sun2014} we have determined the rate of exciton-exciton annihilation in MoS$_2$ to $k_A =$ (4.3 $\pm$ 1.1) $\times$ 10$^{-2}$ cm$^2$/s at room temperature. Using
	\begin{align}
		n(\Delta t) = \frac{n_0}{1+k_An_0\Delta t},
	\end{align}
	with $n_0$ the initial charge carrier density, $n$ at $\Delta t = $1\,ps is estimated. We note that for $n_0 = 1 \times 10^{12}$ cm$^{-2}$ the effective lifetime of the exciton is 11\,ps. 
	
	\subsection{Simulation of HHG process}
	The nonlinear electron-hole dynamics in monolayer TMDCs and the resulting high harmonics are simulated by solving the semiconductor Bloch equations (SBE) in the Houston basis\,\cite{Li2019a}. We use a model Hamiltonian based on a nearest-neighbor tight-binding model for gapped graphene
	\begin{align}
		H=
		\begin{pmatrix}
			\Delta/2 & \gamma f(\textbf{k}) \\
			\gamma f^\ast(\textbf{k}) & -\Delta/2 \\
		\end{pmatrix},
		\label{eq. Hamiltonian}
	\end{align}
	where $\Delta =1.8$\,eV is the band gap, $ f(\textbf{k}) = \exp\left(i\frac{ak_y}{\sqrt{3}}\right) + 2 \exp\left(-i\frac{ak_y}{2\sqrt{3}}\right)\cos\left(\frac{ak_x}{2}\right)$, $\gamma = -3.03$\,eV the hopping parameter between nearest neighboring atoms and $a = 0.315$\,nm the lattice constant. The model band structure consists of two bands and takes the distinct electron dynamics around K and K' into account.
	
	\begin{align}
		i\partial_t \rho_{nm}^{\textbf{k}(t)} = \left[\varepsilon_m^{\textbf{k}(t)} - \varepsilon_n^{\textbf{k}(t)} - \frac{i(1-\delta_{nm})}{T_2}\right] \rho_{nm}^{\textbf{k}(t)} \\
		- \textbf{E}(t) \sum_{m'} \left[\textbf{d}_{m'n}^{\textbf{k}(t)} \rho_{m'n}^{\textbf{k}(t)} - [\textbf{d}_{mm'}^{\textbf{k}(t)} \rho_{nm'}^{\textbf{k}(t)}\right].\nonumber
	\end{align}
	
	Here, $\varepsilon_{n,m}^{\textbf{k}}$ is the energy of the eigenstate in the Houston basis, with $n,m $ denoting the valence and conduction band states. $\textbf{d}_{mn}^{\textbf{k}}$ is the complex dipole matrix element \cite{Motlagh2019}. $T_2$ is a phenomenologically introduced dephasing time, causing an exponential decay of the inter-band polarization, which is proportional to the excursion time of the e-h pair. \\
	We define a linearly $x$-polarized electric field waveform $\text{E}(t)~=~E_{x,0} \exp{(-2\ln(2)(t/\tau_\text{p})^2)}\cos(\omega t )$, with $E_0 = 0.6$\,V/nm the peak electric field strength, $\tau_\text{p} = 100$\,fs the pulse duration and $\hbar\omega = 0.248$\,eV the central photon energy, matching the experimental conditions. Using the Bloch acceleration theorem $k(t) = \hbar^{-1}eA(t)$, the transient dynamic of the electron wavenumber is calculated. $A(t) = -\int_{-\infty}^t E(t')dt'$ is the vector potential. The SBE is solved by applying fourth-order Runge-Kutta method.\\
	
	From the electron dynamics the intra-band current and the inter-band polarization are calculated
	\begin{align}
		j_\text{intra}(t) &= - \sum_m \int_\text{BZ} d\textbf{k}^2p_{mm}(k(t))\rho_{mm}^{\textbf{k}(t)}\\
		p_\text{inter}(t) &= - \sum_{\substack{mm' \\ m\neq m'}} \int_\text{BZ} d\textbf{k}^2p_{m'm}(k(t))\rho_{mm'}^{\textbf{k}(t)},
	\end{align}
	with 
	\begin{align}
		p_{mm}(k) = \nabla_\textbf{k} \varepsilon_m(\textbf{k})
	\end{align}
	and
	\begin{align}
		p_{mm'}(k) = i (\varepsilon_m(\textbf{k}) - \varepsilon_{m'}(\textbf{k})) d_{mm'}(\textbf{k}). 
	\end{align}
	
	\begin{figure}[t!]
		\begin{center}
			\includegraphics[width=8cm]{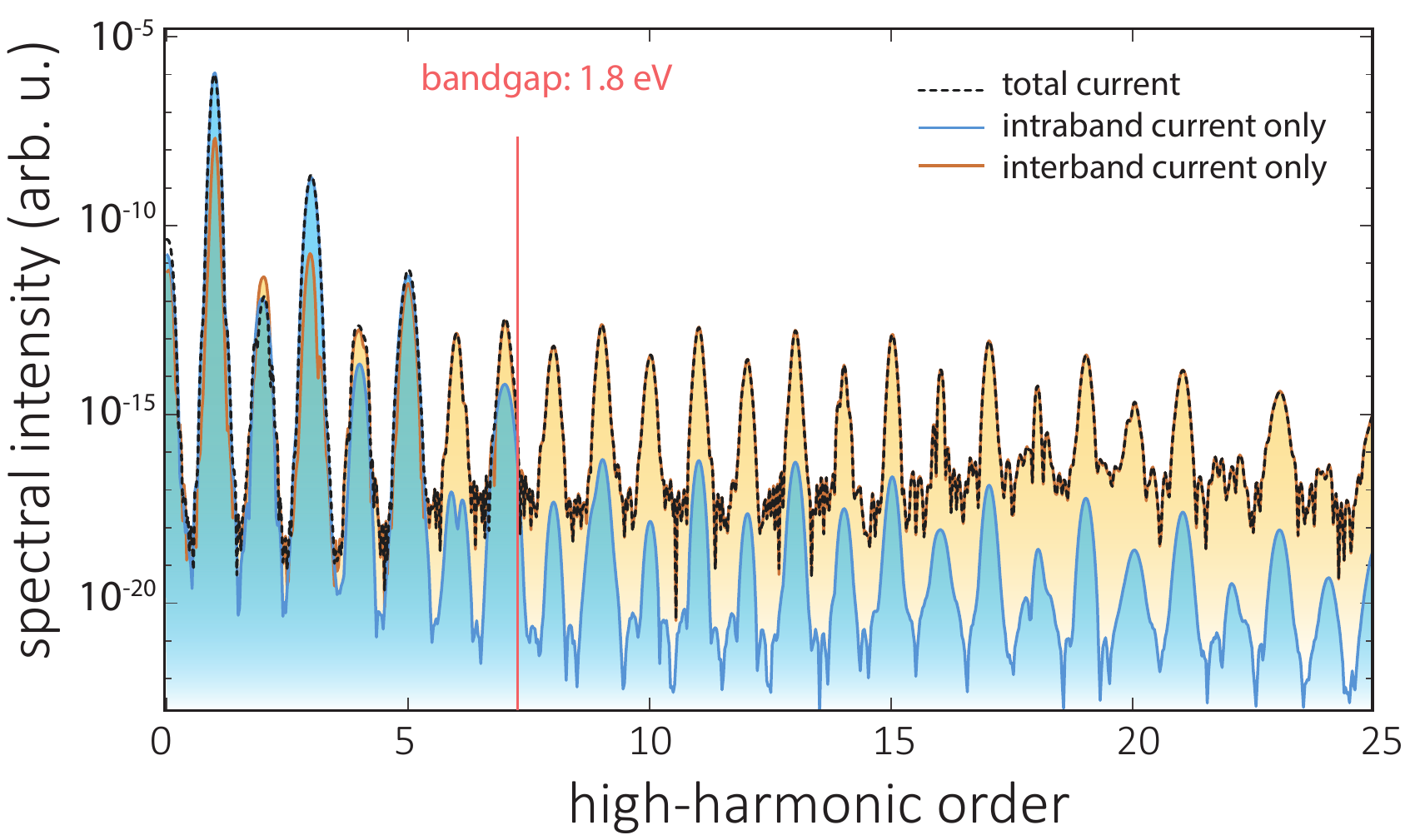}
			\caption{\textbf{Simulated HHG spectrum.} 
				Using the tight-binding model Hamiltonian (Eq.\,\eqref{eq. Hamiltonian}), representing MoS$_2$, the electron dynamics is calculated in the Houston basis. The corresponding HH spectrum for the intra-band current (blue), the inter-band polarization (orange) and the total HHG spectrum is shown (dashed black line). }	
			\label{fig: S3}
			\vspace{-10pt}
		\end{center}
	\end{figure}
	
	By taking the modulus square of the temporal Fourier transform of the sum of the intra-band current and inter-band polarization the high-harmonic spectrum is calculated. Figure~\ref{fig: S3} shows the spectrum for the inter-band polarization (orange), intra-band current (blue) and the total HH spectrum (dashed black line). For low order harmonics, i.e., 2$^\text{nd}$ and 4$^\text{th}$ harmonic order, the intra-band current dominates the HHG response. For high harmonics larger than the 5$^\text{th}$ harmonic order, as measured in the experiment, inter-band polarization is the dominant source of the HHG process. 
	
	\subsection{Analytic model}
	The experimental observation can also be captured by a simple analytical model, which relays on the exponential decay of the inter-band polarization caused by electron-hole dephasing. The intensity of each harmonic order can be described as 
	\begin{align}
		I_\text{HH}(n) \propto \exp \left( - \left( \frac{2\tau_\text{HH}}{T_2(n)} \right) \right),
	\end{align}
	where $\tau_\text{HH}{T_2(n)}$ is the emission time of the respective harmonic order. Similar to the measurement, we define the ratio $\mathcal{R}(n)$ between pump-on, i.e., $T_2(n)$ and pump-off i.e., $T_2(n_0)$
	\begin{align}
		\mathcal{R}(n) = \exp \left( -2\tau_\text{HH} (\frac{1}{T_2(n_0)} - \frac{1}{T_2(n)} ) \right).
		\label{eq. 1}
	\end{align}
	To further simplify the analysis, we linearize $\tau_\text{HH}$ and use Eq.\,\eqref{eq. 1} to qualitatively explain qualitatively the harmonic order dependence (dashed lines in Fig.~\ref{fig: 2}\,\textbf{b} and Fig.~\ref{fig: 3}\,\textbf{f}). We note that the linearization is appropriate for higher harmonics, but breaks down for orders around the bandgap, i.e., the 5th order. We further note that the analytic model is not used for a quantitative analysis. 
	
	\subsection{Drude model}
	The solid line in Fig.~\ref{fig: 3}\,\textbf{g} indicates the dependence of the electron-hole dephasing on the carrier density predicted by a kinetic gas (Drude) model. If screening between the carriers is not considered, then the carrier-carrier scattering time (i.e., the electron-hole dephasing time) will scale $T_2^{-1} \propto n$, i.e. $T(n+n_0)/T(n_0) = n_0/(n+n_0)$, for the normalized pump-probe data, matching our experimental observations \cite{Bigot1991}.
	\subsection{Results for bulk MoS$_2$ samples}
\begin{figure}[h!]
		\begin{center}
			\includegraphics[width=9cm]{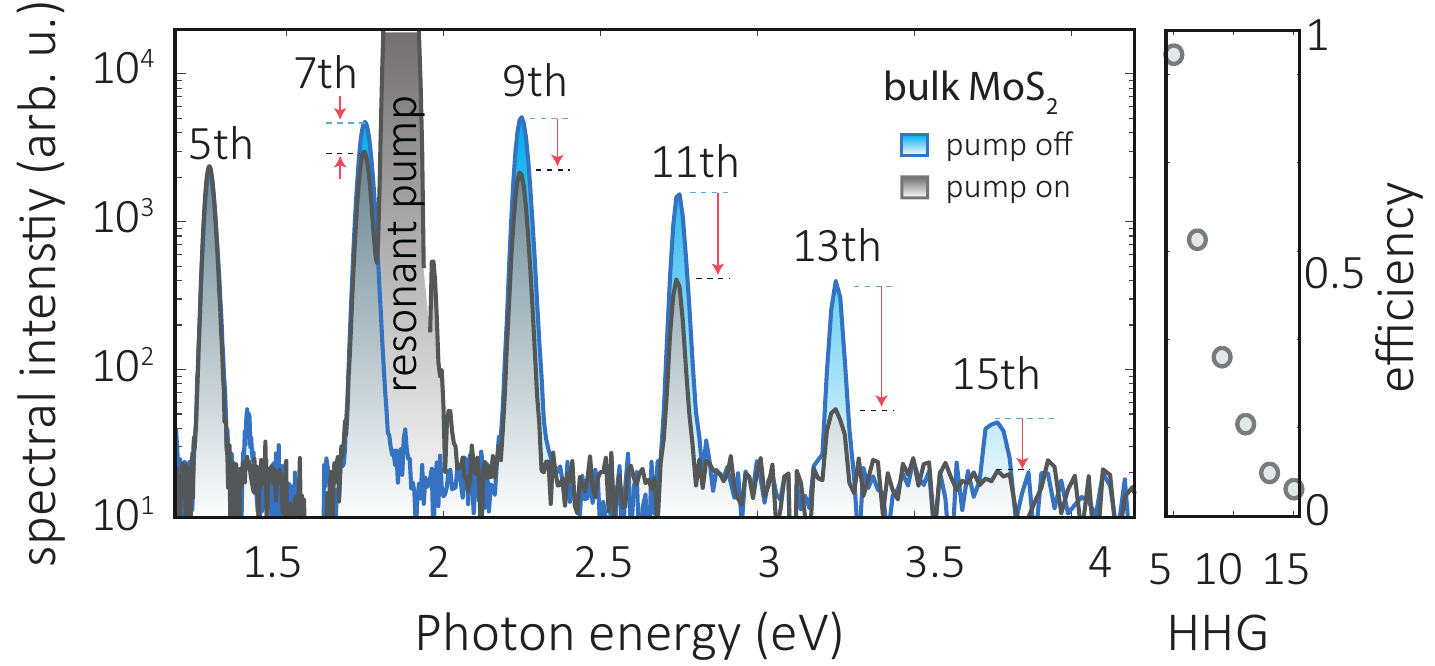}
			\caption{\textbf{Bulk MoS$_2$.} 
				Measured HH spectrum, ranging from 5$^\text{th}$ to 15$^\text{th}$ order for two cases: gray with photodoping and blue: without photodoping. The red arrows compare the peak spectral intensity of each harmonic order for both cases. The integrated and normalized (pump-on/pump-off) efficiency for each harmonic order is shown on the right panel.}	
			\label{fig: S1}
			\vspace{-10pt}
		\end{center}
	\end{figure}

	Bulk MoS$_2$ is prepared from a single crystal bulk from SPI supplies.
	Figure~\ref{fig: S1} shows the HHG spectrum with (gray) and without (blue) additional photodoping. Compared to monolayer MoS$_2$, with a broken inversion symmetry, only odd-order harmonics are observed. Similarly to the monolayer sample, bulk shows a monotonic increase of the HHG yield as a function of the harmonic order (see right panel). 
	
\end{document}